# Document clustering using graph based document representation with constraints

Farnaz Amin
Masters Student of National University of Computer & Emerging Sciences, Karachi, Pakistan

Muhammad Rafi
Faculty of Computer Science, National University of Computer & Emerging Sciences, Pakistan

Mohammad Shahid Shaikh
Faculty of Electrical Engineering, Habib University, Karachi, Pakistan

**Abstract:** *Document clustering is an unsupervised approach in which a large collection of documents (corpus) is subdivided into smaller, meaningful, identifiable, and verifiable sub-groups (clusters). Meaningful representation of documents and implicitly identifying the patterns, on which this separation is performed, is the challenging part of document clustering. We have proposed a document clustering technique using graph based document representation with constraints. A graph data structure can easily capture the non-linear relationships of nodes, document contains various feature terms that can be non-linearly connected hence a graph can easily represents this information. Constrains, are explicit conditions for document clustering where background knowledge is use to set the direction for Linking or Not-Linking a set of documents for a target clusters, thus guiding the clustering process. We deemed clustering is an ill-define problem, there can be many clustering results. Background knowledge can be used to drive the clustering algorithm in the right direction. We have proposed three different types of constraints, Instance level, corpus level and cluster level constraints. A new algorithm ConstrainedHAC is also proposed which will incorporate Instance level constraints as prior knowledge; it will guide the clustering process leading to better results. Extensive set of experiments have been performed on both synthetic and standard document clustering datasets, results are compared with recently proposed methods on standard clustering measures like: purity, entropy and F-measure. Results clearly establish that our proposed approach leads to improvement in cluster quality.*

## 1. Introduction

Document clustering is an unsupervised approach in which a large collection of documents (corpus) is partitioned into smaller and meaningful sub-groups (clusters), concurrently achieving high intra-similarity and low inter-similarity. There are many applications of document clustering in field of information, science, library and business, but the most prominent application where clustering is used is summarized below:

*Organizing Large document collection:* When we query on any search engines, we are displayed hundreds of pages, in which some pages are related to our query and some are not. They are not categorized, making it difficult to identify relevant information. In this situation clustering mechanism can be used to automatically group the retrieved documents in response to our query into a list of meaningful categories, there is one open source software search engine, Carrot2 [1] which do the same. It returns document list divided into different categories, and user can select the relevant category for information retrieval.

*For example*
If we search "Java", Carrot2 categorize the results of a search into groups like "Java developer", "Java Programming," and "Java download" etc.

Clustering is believed to be an indefinite problem, as it can lead to many clustering results. Below example explains this more precisely. Suppose there is a document corpus of 4 documents Doc = {D1, D2, D3, D4}, after doing document clustering following different cluster arrangements could be produced:

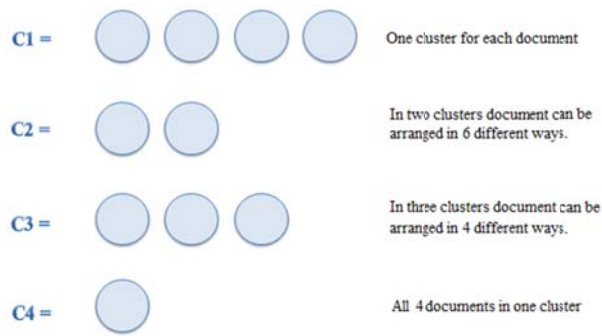

**Figure 1: Different Clustering Results**

In order to deal with this problem, many researchers have proposed various techniques for improving the document representation and produce semantic grouping of data and also suggested various approaches in the area of semi-supervised learning which includes use of constraints for guiding the clustering process for the identification of meaningful clustering result.

There are numerous issues related to document clustering which are necessary to be dealt with. There are three main components of a document clustering approach, these are (i) How to represent the document, as it captures the semantics of document, and also contribute to reduce the problem of high dimensionality, (ii) define a similarity measure, for computing the pair-wise similarity of documents, and which is able to assign high values to the documents which are semantically similar, and (iii) which clustering algorithm is used to finally cluster the documents.

For meaningful document representation, lot of efforts has been made. The most popular and basic model of document representation is the vector space model where each document is referred as "Bag of words" and this technique is synonymously termed as BOW technique, which does not consider association between words. In order to address the semantic and meaningful representation some techniques also represent document as concept vector. It utilizes Wikipedia for mapping document terms to concepts in Wikipedia. There are number of approaches that use graph-based representation of documents, which depict the dependencies of words, explaining the association between words of documents. Graph based technique addresses the challenge of considering the natural language relationship between the words which was ignored by other proposed techniques.

There are different clustering algorithms which groups the data into different categories, but it is important that the results obtained after clustering correspond to the true user desired semantic grouping of data. Traditional clustering algorithm due to its unsupervised nature are unable to provide meaningful and desired clustering result, for example, many document talking about different topics but having similar words will be grouped together. Solution of this problem is to introduce background knowledge. This can be provided in form of constraints. Constraints are kind of supervisory information, where domain knowledge is utilized for guiding the process of document to assign to target clusters. Algorithms are proposed which incorporate constraints making clustering process a semi-supervised approach.

To deal with the above discussed challenges of document clustering we have proposed technique which will use graph based document representation with the use of constraints (DCGBDRC). Our goal revolves around the idea of providing a meaningful document representation and semantic user-desired grouping of data. A graph data structure can easily capture the non-linear relationships of nodes, document contains various feature terms that can be non-linearly connected hence a graph can easily represent this information. We have used approach of Wang, et al. [2] which represent document as graph and then extended this approach by suggesting different ways of integrating background knowledge into clustering algorithm. Our approach is therefore different from traditional unsupervised clustering.

We are particularly interested in incorporating three different kinds of constraints.

*Instance level constraints*
Instance level constraints consist of Must-link (ML) and Cannot-link (CL) constraint.

- Must-link constraints state that two instances must belong to the same cluster.
- Cannot-link constraints specify that two instances must not be placed in the same cluster.

*Corpus level constraints*
Corpus level constraints explains having prior knowledge about the datasets containing data instances sharing same information, and so it is termed as corpus/dataset level constraints. Documents belonging to same dataset/newsgroup should be placed in one cluster.

*Cluster level constraints*
Cluster level constraint refers to two different types of constraints.

- *Size constraints:*

Restriction on number of documents assigned to each cluster.

- *Existence of sub-graph in clustered documents:* Documents sharing same sub-graph should be clustered together. Same sub-graph shows that any two documents share same information about the topic.

The major contribution of our work is the idea of integrating prior knowledge in form of constraints, and for this we have developed an effective and efficient *ConstrainedHAC* algorithm, modification to the existing traditional HAC algorithm, which will now deal with constraints. Experimental results are evaluated on both, benchmark real-world datasets and synthetic data sets that illustrate the performance of our proposed semi-supervised approach, C*onstrainedHAC* algorithm and its comparison with traditional HAC algorithm.

The rest of the paper is organized as follows: Section 2 reviews the related work. In section 3, we describe our proposed approach. Section 4 presents our experiments. Section 5 discusses the results, Finally, Section 6 summarizes our work and Section 8 elaborates our future research work.

## 2. Literature Review

Different approaches have been adopted for document clustering, focusing on how to represent the document and improve the clustering result. Semi-supervised learning based approaches are also proved to be very effective that suggests intervention of constraints for guiding clustering process. We have divided the related work in two main broad categories i.e. document representation and constrained clustering. In this section, we briefly summarize the work in these areas.

**2.1 Document Representation**

Document clustering method generally comprises of three steps, (i) Document representation, (ii) similarity measure and (iii) actual clustering algorithm. Representations of document signify that finding a document model, a set of features that can be used to represent a document.

The most common used model of document representation is the Vector space model (VSM) referred as bag of words, in which the document is converted into a vector of words having no relationship between words. To cater the relationship of words Document can also be represented as Concept vector [3]. In this model document is regarded as "Bag of concept" (BOC) each concept having weight. In this Wikipedia is utilized for mapping document terms to concepts in Wikipedia. Semantic connection among concepts is incorporated in "document similarity measure". Some research work used graph to provide meaningful document representation [4], [5], [2] capturing the word relationship. In [4], document graph representation technique GDCLUST is proposed which converts the whole document into document graph. GDClust is different from other existing clustering techniques because it is able to assign documents in the same cluster even if they do not contain common keywords, but still have same sense, as it looks at the origin of the keyword in document graph. Document is converted to its document graph using document graph construction algorithm which utilizes BOW toolkit and WordNet 2.1 taxonomy. The Topic Map model [6] is one of those models which capture semantic content of the document. Topic map data structure is very similar to concept graph; it does not only represent the topics present in the document but also captures the occurrence and association between documents. They have proposed a similarity Measure to check the relatedness between pair of documents, which calculates the no of common topic, common topic tags and tag value between documents. Hierarchical agglomerative clustering is used to perform the clustering.

Recently few techniques were suggested which focus on the natural language relationship between words. Wang, et al. [2] technique is one of those approaches. They have provided a graph based document representation in which documents are represented as dependency graph. Nodes are characterized as words (which can be seen as Meta-description of document) and edges represent the relationship between pair of words. In DG model document A and document B corresponds in any to the same dependency graph indicating that they are semantically equal with each other. As it captures the semantic content of document providing a meaningful representation, we have used this model for document representation. The similarity measure is the basic component of any technique. The authors have suggested a similarity measure in which pair wise similarity of documents is computed based on their corresponding dependency graphs; we have also used the same measure to compute the similarity. Moreover Theodosiou, et al. [5] also proposed a document clustering technique for biomedical dataset. This retrieves relevant information from biomedical repository. This novel approach represent document as vector of weighted words also known as VSM (vector space model). It also retrieves its relevant documents from PubMed. The novelty lies in the idea of representing the relationship between the documents with association graph, where each vertex represent a document and edges represent document relatedness based on the related document information from PubMed. Documents are clustered using Markov clustering algorithm (MCL), an unsupervised clustering algorithm for graph.

The third major step is the right choice of clustering algorithm. The most commonly used algorithm is agglomerative and hierarchical clustering algorithm. K-means is another popular clustering algorithm that has been used in a variety of application domains. We have also used group average hierarchical clustering algorithm and modified it to produce new *constrainedHAC* algorithm to deal with constraints.

**2.2 Constrained Clustering**

Lately few researchers have made an effort in the area of semi supervised learning approaches, which showed effective results. Prior knowledge in form of constraints proved to produce better clustering results as compared to traditional un-supervised clustering. Constraints are a common way to add background knowledge to the clustering algorithm, advising that which data points (documents) should be clustered together or not. Constrained clustering mostly use instance level constraints such as "must-link" and "cannot-link" to guide the unsupervised clustering [7], [8], [9], [10]. K. Reddy and Anand [11] proposed an algorithm to systematically add instance-level constraints, which enforce constraints on data points, to graph based clustering algorithm CHAMELEON. Proposed algorithm, Constrained CHAMELEON (CC), embed constraint in first phase of algorithm. Constraints (must link (ML) and cannot link (CL)) were added before graph partitioning. They selected the best results obtained by chameleon algorithm and showed that these results can be improved by adding constraints.

Constraints are not limited to 1D clustering but efforts have been made in coclustering also where both document and word relationship are studied at the same time compared to traditional clustering which focus on only document relationship. Methods have been proposed for extending coclustering to semi-supervised coclustering by incorporating both supervised and unsupervised word and document constraints [7]. Supervised constraints includes human annotated categories, whereas unsupervised constraints include automatically constructed document constraints based on the overlapping named entities by an NE extractor. If there were some overlapping NEs in two, then a must-link can be created between those two documents. Song, et al in [7] have discussed the overall effectiveness of both types of constraints and evaluated the results.

Traditional pairwise constraints obtained from human experts may conflict with each other and they are not always correct, techniques are suggested how to remove noise from those pairwise constraints. A new concept of **Elite pairwise constraints** is proposed by Jiang, et al. in [9]. In this authors have taken a step by introducing such constraints which will not conflict with each other and will guide the clustering process in the right direction. They have also discussed that it is NP-hard to acquire Elite pairwise constraints but used Limit crossing heuristic algorithm to extract some part of these constraints.

On which level constraints should be incorporated, where these constraint will produce better and effective results, is a big question. In [11], K. Reddy et al. embedded constraints into the clustering algorithm CHAMELEON through learning a new distance (or dissimilarity) function, while authors in [8] authors have discouraged this technique giving reason that as clustering is the task of dividing the collection into meaningful clusters, so pairwise constraints should be employed during the clustering process rather than modelling the similarity matrix with these constraints and then perform clustering and for the same we have also incorporated constraints during clustering process by modifying the existing HAC algorithm.

## 3. Proposed Approach (DGBC)

The proposed approach for document clustering is named as "Document clustering using a graph based document representation with constraints". (DGBC)"

**3.1 Graph based document Representation.**

We have represented document as a document graph. Graph data structures can easily captures the non-linear relationships of nodes, the documents contains various feature terms that can be non-linearly connected hence a graph can easily represent this information.

Document is represented as dependency graph approach presented by Wang, et al. [2]. Document dependency graph $G$ is denoted as $G = (V,E)$ where $V = \{v_1, v_2, ..., v_n\}$ is the set of vertices in the graph, each vertex $v_i$ represent word $w_i$ of document and $E =$

*{e1, e2, ..., em}* is the set of edges, where each edge *ej* between vertices indicate that there is some relationship between words of document.

Consider below example having document 1 and 2.

**Doc 1:** Maker of iPhone is "APPLE". Steve Jobs was the CEO at Apple.

**Doc 2:** The CEO of Apple was Steve jobs. "IPHONE" maker is Apple.

Dependency graph of document 1 is shown in figure 2.

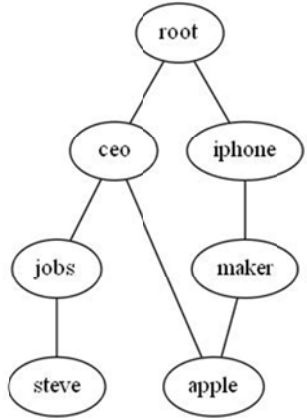

**Figure 2: Document Graph for ": Maker of iPhone is "APPLE". Steve Jobs, was the CEO at Apple"**

Both documents have different words sequence but they should have high similarity as they are semantically same, talking about same topic. In the model presented by Wang, et al. in [2] both documents, document *1* and 2 correspond to the same dependency graph which correctly indicates that they are semantically equal with each other.

In practice, document graph construction is done in following steps: (1) Data cleaning is performed by removing all the numbers and delimiters, eliminating triple and double spaces and converting all the words into lower case (2) Document is divided into sentences and then stop words are removed from the sentences (3) Stanford parser [12] is used to obtain word dependencies from the cleaned sentences. (4) Using word dependencies, obtained from Stanford parser, and non-stop words of document, the graph is constructed in following steps:
 a. Vertices and edges are added by processing each sentence in the document.
 b. For each sentence, we parse it using the dependency parser, which outputs a set of words and the identified pairwise relations between them.
 c. Pairwise relation between words represents an edge between vertices and non-stop words represent set of vertices.
 d. The lengths of all the edges in the graph are set to 1.

(5) Obtained dependency graph is converted into a weighted dependency graph by calculating weight of every vertex using *tf-idf* measure. Each vertex $v_i$ in graph G correspond to word $w_i$ of document. (6) Calculate similarity between document graphs. Document graph is converted into feature weight matrix. Similarity measure of Wang, et al. [2] is used to calculate similarity between two document graphs.

### 3.2 Constrained Clustering

Another component of our proposed approach is introducing additional background knowledge in form of constraints which will facilitate the document clustering process. We have proposed three different kinds of constraints, which can be incorporated in the clustering algorithm to guide the clustering towards better, improved and meaningful clusters. In the next section we have discussed the types of constraints we are using, and then we have discussed our proposed *ConstrainedHAC* algorithm.

#### 3.2.1 Constraints

We are particularly interested in incorporating three different kinds of constraints. These are discussed below.

*Instance level constraints*
Supervisory information in form of Instance level constraints includes "Pairwise constraints", most popularly used in semi-supervised clustering. Pairwise constraints define the relationship between any two data instance, whether they belong to same cluster or not.

Pairwise constraint specifically consists of two kinds of constraints.
- **Must-link (ML):** It indicates that two data instances must be assigned to the same cluster
- **Cannot-link (CL):** It indicates that two data instances must be assigned to different clusters

*Example:*
Document corpus: {d1, d2, d3}
Must-link (ML): {d1, d2}
Cannot-link (CL): {d1, d3}

d1 = "Steve Jobs was the CEO at Apple".

d2 = "Steven Paul Jobs was the co-founder, chairman, and CEO of Apple Inc. Jobs also co-founded and served as chief executive of Pixar Animation Studios".
d3 = "I'm eating the most delicious apple".

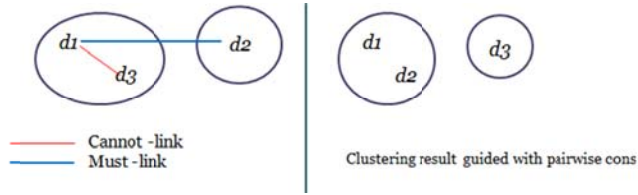

**Figure 3: Guided Clustering after applying ML and CL constraints**

*Corpus level constraints*

Corpus level constraints indicate supervisory information about the datasets, containing data instances sharing same information, and so it is termed as corpus/dataset level constraints.

For e.g. Dataset collection of news feed of particular event. For e.g. **14th November, 2013, Sachin Tendulkar retirement from test cricket** or **Benazir Bhutto assassination on 27th December 2007**. Every news will be conversing on this topic.

As our Document Representation is in form of graph, so there will be a common sub-graph, depicting the similarity among all news feed in the corpus, thus guiding the clustering process and forcing these documents to be assigned to same cluster.

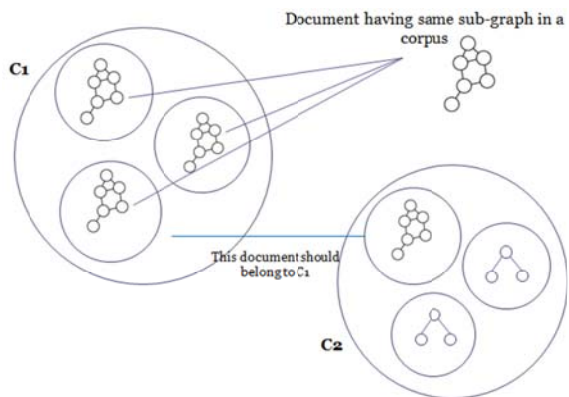

**Figure 4: Guided Clustering with Corpus level constraints**

*Cluster level constraints*

Cluster level constraint refers to two different types of constraints

*Size constraints*
- Impose restriction on number of documents assigned to each partition.
- Based on prior information about the data collection, size of clusters are specified

*Existence of sub-graph*
- Documents clustered together should all share a common sub-graph (known from corpus level constraint).
- If there occurs an existence of the same sub-graph in a document belonging to different cluster, then it will be a constraint to move that document to the cluster having similar documents.

*Example*
Document corpus: {d1, d2, d3, d4, d5, d6, d7, d8, d9, d10}
Size Constraint: {no1, no2, no3} = {5, 3, 2}
Where "no" being the number of objects in each cluster.

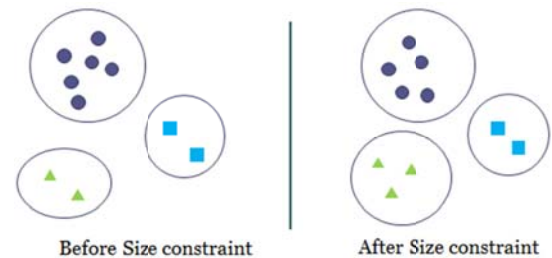

**Figure 5: Guided Clustering after applying Clsuter level size constraints**

We have incorporated instance level constraints in hierarchical Agglomerative clustering and devised ConstrainedHAC algorithm dicussed in next section.

### 3.2.2 Constrained Algorithm

To incorporate instance level constraints in hierarchical Agglomerative clustering we have modified the algorithm. Constrained Hierarchical Agglomerative clustering (*ConstraintedHAC*) with must-link and cannot-link constraint as shown in table 1.

ML and CL constraints constructed are humanly annotated. There exists transitivity of constraints. When constructing set of ML and CL constraints, we take transitive closure[1] over the constraints and then set of derived constraints is used in algorithm.

1. Construct cannot link and must link constraints.
2. Take a transitive closure over the constraints.
3. Compute the Similarity matrix between the data points (documents).
4. ObjectLabel ← [ ] (keeps track of cluster).
5. Let each data point be a cluster.

---

[1] *The closure is performed over both kinds of constraints. for e.g., if di must link to dj which cannot link to $d_k$, then di cannot link to $d_k$.*

6. *Apply_ML_Constraints(Similarity Matrix); // Apply ML constraints initially and merge all documents which have a must link constraint between them.*
7. **Repeat**
8. Start by selecting two data points having max similarity. (It should be non- negative).
9. For the two closest clusters, Use *Validate_CL_constraint ( );// It will validate that two cluster being merged are not violating any CL constraint, if they are violating CL constraint then don't merge and update the similarity matrix by updating similarity of $d_i$ and $d_j$ to -1, else* **Merge ( );**
   If Merge ( ) then Boolean Change_in_cluster = true
   Else Change_in_cluster = false

10. Objectlabel.Append (store merging of cluster).
11. Update the similarity matrix.
**Until** *No_chage_in_cluster*.

**Table 1: ConstrainedHAC Algorithm**

Initially all documents are in their own cluster same as HAC algorithm, then Must link constraints are applied, merging the documents in a cluster which fulfil a must-link constraint. For e,g if $d_i$, $d_j$ have a ML constraint then both will be assigned to same cluster, similarly if $d_k$ and $d_l$ have ML constraints, they will be merged together in one cluster. After applying ML constraint HAC clustering starts by selecting any two clusters for merging which have the maximum non-negative similarity. For the two closest clusters to be merged, CL (cannot-link) constraints are validated, which ensures that two clusters being merged are not violating any CL constraints. If they are violating any CL constraint then clusters are not merged together and similarity matrix is updated by assigning penalty of -1 to similarity of $d_i$ and $d_j$, else the two clusters are merged together.

Hierarchical Agglomerative clustering increasingly join two closest clusters by reducing the number of clusters by 1 i.e. the stopping condition of HAC is when the number of cluster reaches 1. We have modified the stopping condition by keeping track of the change in cluster. As we are dealing with ML and CL constraint and we stop the two clusters being merged if they are violating any CL constraints, so a point will come where there will no clusters left to be merged leaving the state of cluster unchanged. Thus our clustering algorithm will stop and outputs the partition satisfying all constraints when no change in cluster will be observed.

Figure 6 shows all steps involved in document clustering using graph based document representation with constraints.

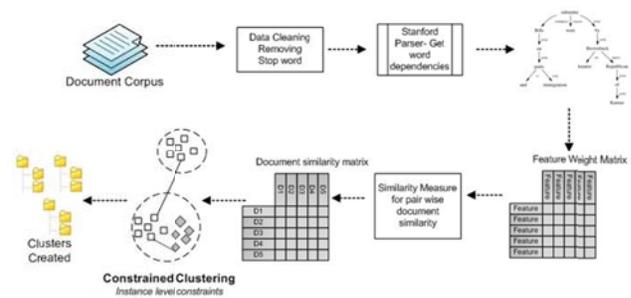

**Figure 6: Steps involved in Document clustering using graph based document representation with constraints.**

## 4. Experimental Setup

The algorithmic method proposed in this paper, is implemented and tested through a series of experiments. Our approach is implemented in Java programming language. The experiment is executed on a HP Pavilion g6 Notebook, with Intel Core i5 processor and 4GB of RAM with 500GB of Hard Disk storage.

### 3.1 Datasets

**Standard dataset**

Three standard datasets *NEWS20, Reuters* and OHSUMED are considered for evaluating the clustering method.

These data sets are selected for the reason because these are commonly used by researchers for conducting experiments.

- *NEWS20:* A popular data set used for experiments in text applications of machine learning techniques is what's known as the 20 newsgroups collection. These techniques include text classification and text clustering. The dataset is available at http://qwone.com/~jason/20Newsgroups/
- *Reuters:* It is the most common dataset used for evaluation of document categorization and clustering. This is a collection of documents that appeared on Reuters newswire in 1987. The documents were assembled and indexed with categories. The dataset is available at http://www.daviddlewis.com/resources/testcollections/reuters21578/
- *OHSUMED:* The OHSUMED dataset is a subset of the MEDLINE database. It is the on-line bibliographic medical information database maintained by the National Library of Medicine. The dataset collection consists of titles and abstracts from 270 medical journals abstracts over a five-year period (1987-1991). The data set is

available at
http://disi.unitn.it/moschitti/corpora.htm

**Synthetic datasets**

For the purpose of conducting experiments after incorporating constraints and evaluation of result we have created a dataset which contains news document. News documents are related to different Microsoft products.

We have selected subset of these datasets for evaluation of results.

**Table 2: Experimental Datasets**

| Data Sets | Data Source | No. of Doc |
|---|---|---|
| D1 | NEWS20 | 100 |
| D2 | NEWS20 | 200 |
| D3 | NEWS20 | 400 |
| D4 | Reuters | 100 |
| D5 | Reuters | 200 |
| D6 | Reuters | 400 |
| D7 | OHSUMED | 100 |
| D8 | OHSUMED | 200 |
| D9 | OHSUMED | 400 |
| D20 | Synthetic Dataset | 100 |

### 3.2 Background knowledge as Constraints

After reading documents from both standard and synthetic datasets, Pair-wise constraints Must-link (ML) and cannot-link (CL) are manually identified. All constraints are provided by human and we have utilized these in our algorithm to conduct experiments.

### 3.3 Evaluation Measures

Effectiveness of cluster quality can be measured using different evaluation measures; we validate the effectiveness of our proposed approach by using standard cluster quality measures like F-Measure, Purity and Entropy.

*3.3.1 F-Measure*

The F-measure tries to capture how well the groups of the investigated partition at the best match the desired set of classes

Consider the resulting cluster as j and the desired set of documents as class i. F-measure used both recall and precision for calculation. Recall and precision of cluster j with respect to class i is calculated as follows:

$$prec(i, j) = \frac{c_{ij}}{c_j}$$

$$rec(i, j) = \frac{c_{ij}}{n_i}$$

Then F-measure of cluster j and class i is defined as follows:

$$F_{ij} = \frac{2 * recall(i, j) + prec(i, j)}{prec(i, j) + recall(i, j)}$$

F measure of overall clustering can be calculated as:

$$F = \sum_i \frac{n_i}{n} \max(F_{ij})$$

Where n denotes the number of documents in a dataset and $n_i$ represents the number of document in cluster i. F-score value will be between 0 and 1, and larger value of f-score indicated higher clustering quality.

*3.3.2 Purity*

Purity is an external evaluation criterion for clustering. Each cluster is assigned to the class which is most frequent in the cluster. Formally purity of cluster j is defined as:

$$purity(j) = \frac{1}{c_{ij}} \max(c_{ij})$$

Purity of entire clustering is the weighted sum of the individual clustering purities. It can be computed as:

$$purity = \sum_j \frac{c_{ij}}{N} purity(j)$$

*3.3.3 Entropy*

Entropy measures how the various semantic classes are distributed within each cluster. Each cluster j should be homogeneous, that is, the class distribution within each cluster should tend to a single class, which is zero entropy, and smallest possible value for entropy.

Smaller entropy values indicate better clustering quality; signifying less disorder in a clustering,

It can be computed as:

$$E_j = -\sum_i p_{ij} \log p_{ij}$$

The total entropy is calculated as the sum of the entropies of each cluster weighted by the size of each cluster:

$$E_{en} = \sum_{j=1}^{m} \frac{n_j * E_j}{n}$$

## 5. Results and Discussion

We proposed a method for incorporating background knowledge to improve clustering process. We utilized background knowledge in form of two types of instance level constraints, must-link and cannot-link. We modified the existing Hierarchical agglomerative clustering algorithm and proposed *ConstrainedHAC* algorithm to incorporate these constraints.

Experimental results show that our technique produced significant results in terms of quality of clusters. Document clustering using graph based representation with constraints performed better when compared with the results when no constraints were applied. Experimental test were performed on both synthetic and well-known standard datasets. Purity, Entropy and F-measure of the implemented technique with constraints and without constraints is shown in Table 3 and Table 4. Increased value of Purity clearly indicates that good clusters are produced. Entropy is another measure; its value indicates the disordering in clusters. Reduction in entropy states improvement in results, which signifies less disordering in clusters.

|    | Data Source       | No. of Documents | Purity | Entropy | F-score |
|----|-------------------|------------------|--------|---------|---------|
| D1 | NEWS20            | 100              | 0.66   | 0.27    | 0.65    |
| D2 | NEWS20            | 200              | 0.68   | 0.26    | 0.74    |
| D3 | NEWS20            | 400              | 0.70   | 0.23    | 0.65    |
| D4 | Reuters           | 100              | 0.64   | 0.29    | 0.63    |
| D5 | Reuters           | 200              | 0.66   | 0.27    | 0.75    |
| D6 | Reuters           | 400              | 0.69   | 0.25    | 0.73    |
| D7 | OHSUMED           | 100              | 0.53   | 0.33    | 0.63    |
| D8 | OHSUMED           | 200              | 0.69   | 0.25    | 0.55    |
| D9 | OHSUMED           | 400              | 0.76   | 0.20    | 0.84    |
| D10| Synthetic Dataset | 100              | 0.58   | 0.31    | 0.49    |

Table 3: Purity, Entropy and F-measure of Document clustering Using Graph based representation "without constraints"

|    | Data Source       | No. of Documents | Purity | Entropy | F-score |
|----|-------------------|------------------|--------|---------|---------|
| D1 | NEWS20            | 100              | 0.83   | 0.15    | 0.77    |
| D2 | NEWS20            | 200              | 0.85   | 0.14    | 0.80    |
| D3 | NEWS20            | 400              | 0.89   | 0.10    | 0.86    |
| D4 | Reuters           | 100              | 0.80   | 0.17    | 0.65    |
| D5 | Reuters           | 200              | 0.81   | 0.16    | 0.80    |
| D6 | Reuters           | 400              | 0.84   | 0.14    | 0.86    |
| D7 | OHSUMED           | 100              | 0.84   | 0.14    | 0.87    |
| D8 | OHSUMED           | 200              | 0.89   | 0.10    | 0.86    |
| D9 | OHSUMED           | 400              | 0.92   | 0.07    | 0.84    |
| D10| Synthetic Dataset | 100              | 0.75   | 0.21    | 0.74    |

Table 4: Purity, Entropy and F-measure of Document clustering Using Graph based representation with constraints.

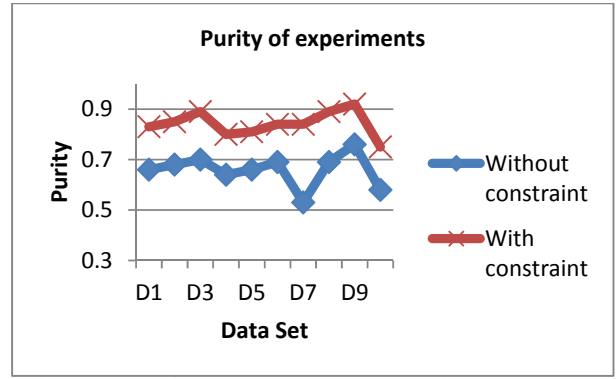

Figure 7: Purity for Document clustering using graph based representation "with constraints" and "without constraints"

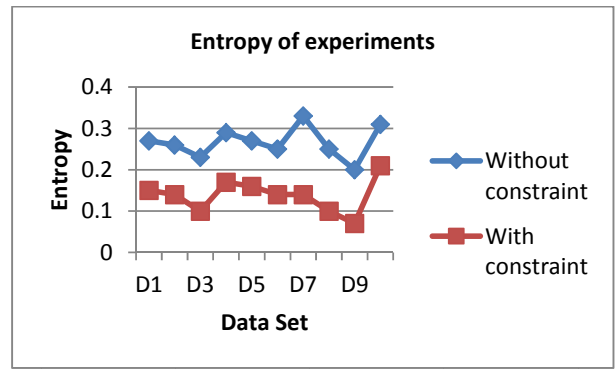

Figure 8: Entropy for Document clustering using graph based representation "with constraints" and "without constraints"

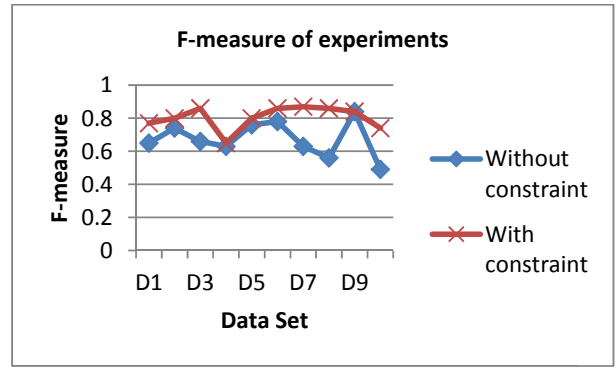

Figure 9: F-measure for Document clustering using graph based representation "with constraints" and "without constraints"

## 6. Conclusion

We have demonstrated that our approach of document graph representation and adding of background knowledge in form of constraints into hierarchical agglomerative clustering produce better clustering results and hence this technique has contributed to this new emerging field of document clustering. We proposed three different types of constraints and devised *ConstrainedHAC* algorithm by incorporating two types of Instance level constraints must link and cannot link, which specified that which two data instances can belong to same cluster and which of them cannot go together in the same clusters. Through our experimental results and evaluation we have proved that addition of instance level constraints improved the quality of clusters produced.

Our experimental results, the values of purity, Entropy and F-score clearly showed that addition of ML and CL constraints in HAC algorithm have significant effect on clustering result and it has greatly improved the overall dendogram formed.

To investigate the effect of number of constraints on clustering performance we varied the number of pair wise constraints by randomly selecting the constraints from the set of human annotated constraints. From the values of purity and entropy we can state that the number of instance level constraints have a significant impact on the clustering performance. As the number of document increases, and more the constraints were added, the better the clustering results were achieved. The variation of number of documents and number of constraints also proved that our proposed *ConstrainedHAC* algorithm is Scalable, which means that value of purity increase with the increment in number of documents and number of constraints. The F-measure for dataset D4, which is a subset of Reuters shows that when small numbers of documents with less number of constraints is used, the result is similar to the non-constrained version. This all shows that our proposed approach (DGBC) has given effective improvement in all test cases and outperformed the non-constrained document clustering.

## 7. Future Work

We have presented an approach to incorporate background knowledge in form of constraint by modifying the existing hierarchical agglomerative clustering algorithm and through experiments we have shown significant improvements.

There are several directions of future work, which includes making use of the other two types of constraints, which are cluster level constraints and corpus level constraints. We intend to extend this algorithm to include these types of constraint as well.

Cluster level constraints refer to size constraint, where we can restrict the number of documents contained in each cluster. This type of constraint can also be incorporated in our proposed *ConstrainedHAC* algorithm, which will have prior knowledge about the size of constraint and it will be utilized by the *Merging* function of *ConstraintedHAC* algorithm. Second type of constraint is the corpus level constraints; it indicates supervisory information about the datasets, containing data instances sharing the same information. Algorithm can be modified in a way to deal with subgraphs. If the two documents share the common subgraph than they will belong to same cluster imposing on them a courpus level constraints because the common subgraph shows that both documents share the same information and they belong to the same dataset. This type of constraint will be of great help as background knowledge; and it will lead to produce better, meaningful and user desired clustering arrangements.